\documentclass{article}

\usepackage[utf8]{inputenc}
\usepackage[T1]{fontenc}
\usepackage{amsmath}
\usepackage{amssymb}
\usepackage{mathpazo} 
\usepackage{helvet} 
\usepackage{booktabs}
\usepackage{natbib}
\usepackage{graphicx}
\usepackage[font=small]{caption}
\usepackage{subcaption}
\usepackage{xcolor}
\usepackage{microtype}
\usepackage{placeins}
\usepackage{float}
\usepackage{hyperref}
\usepackage{url}
\usepackage{doi}

\definecolor{DarkBlue}{rgb}{0.0, 0.15, 0.55}
\definecolor{Gray}{gray}{.25}

\hypersetup{
    colorlinks=true,
    linkcolor=DarkBlue,
    citecolor=DarkBlue,
    urlcolor=DarkBlue,
    pdftitle={Bargmann Zeros as a Diagnostic of the Tunneling Transition in Double-Well Quantum Systems},
    pdfauthor={Tughanbulut Kurtulush, Maciej Janowicz},
    pdfkeywords={Bargmann representation, complex zeros, Husimi function, anharmonic oscillator, double-well tunneling, wavefunction topology},
}

\graphicspath{{figures/}{./}}
\title{Bargmann Zeros as a Diagnostic of the Tunneling Transition in Double-Well Quantum Systems}

\author{
    \textbf{Tughanbulut Kurtulush} \\
    Faculty of Computer Engineering \\
    Vistula University \\
    Warsaw, Poland \\
    and
    \textbf{Maciej Janowicz}\thanks{Corresponding author.} \\
    Institute of Information Technology \\
    Warsaw University of Life Sciences \\
    Warsaw, Poland \\
    \texttt{maciej\_janowicz@sggw.edu.pl} \\
}

\begin{document}
\maketitle

\begin{abstract}
Complex zeros of wavefunctions represented as entire functions in
Bargmann--Fock space encode structural information about the underlying
quantum state. Prior work employed zero galleries of randomly generated
polynomial superpositions of Fock states as visual fingerprints suitable
for classification. Here we examine whether Bargmann zeros of physically
realized eigenstates of one-dimensional anharmonic and double-well
Hamiltonians carry a recognizable signature of the tunneling transition
in the symmetric double well. Ground and first-excited eigenstates are
obtained from a variational ansatz consisting of a physically motivated
symbolic envelope multiplied by a small flexible correction network,
trained by Rayleigh--Ritz minimization of the finite-difference
Hamiltonian expectation value and validated to reproduce energies to
within $\sim 10^{-5}\,\mathrm{Ha}$. The resulting wavefunctions are
projected onto the harmonic-oscillator basis and the complex zeros of the
truncated Bargmann polynomial are located by numerical root-finding.
For harmonic and quartic-anharmonic potentials the zeros show no
preferred orientation. For double-well eigenstates, by contrast, the
zeros condense onto the imaginary axis. A sweep of the barrier parameter
$a$ from $0.5$ to $2.3$ reveals a continuous migration of zeros toward
the imaginary axis, concurrent with the exponential collapse of the
tunneling splitting $\Delta(a) = E_1 - E_0$ over $3.5$ decades.
This condensation is traced to a sign-alternation pattern in the
Fock-coefficient spectrum that is characteristic of bimodally localized
wavefunctions. The complex zero set of the Bargmann-represented
wavefunction thereby provides a compact, purely analytic diagnostic for
the tunneling regime of one-dimensional double-well Hamiltonians,
extending the random-polynomial zero-image framework to physical
eigenstates.
\end{abstract}

\vspace{0.5cm}

\section{Introduction}
\label{sec:intro}

The Bargmann (or Fock--Bargmann) representation maps square-integrable
wavefunctions of a one-dimensional quantum system onto entire
holomorphic functions of a complex variable $z \in \mathbb{C}$, with
Fock states playing the role of the monomial basis $z^n/\sqrt{n!}$
\citep{bargmann1961fock}. This construction sits
within the broader coherent-state framework that originated with
Glauber's quantum-optical coherent states
\citep{glauber1963coherent} and was generalized by Perelomov to
arbitrary Lie groups \citep{perelomov1972coherent}; the standard
reprint volume of Klauder and Skagerstam
\citep{klauder1985coherent} collects the definitive results. Any wavefunction can
therefore be written as a (formally infinite) polynomial in $z$, and
its \emph{complex zeros} form a discrete set in $\mathbb{C}$ that, by
Weierstrass--Hadamard factorization, determines the state up to a
multiplicative entire function of order at most one.

Recent work by Janowicz \& Zembrzuski \citep{janowicz2022bargmann}
explored the visual structure of these zero sets for
\emph{random-coefficient} superpositions of Fock states
and demonstrated that a convolutional neural network can classify
qualitatively distinct families of such random states from their
zero images alone. In parallel, Cerf \emph{et al.}
\citep{cerf2025zeros} have shown that, under a mild energy
condition, a bosonic wavefunction extends to an entire holomorphic
function on $\mathbb{C}$ whose zero set characterizes the
non-Gaussianity of the state via a Hudson-type theorem, and
evolves as a classical many-body dynamical system under Gaussian
Hamiltonians. Together these results raise a natural question:
do the zero images of \emph{physically solved} eigenstates carry
analogous structure, and does that structure encode a recognizable
signature of the underlying physics?

We address this question for the symmetric double well. A
variational ansatz --- a physically motivated symbolic envelope
multiplied by a small neural correction, trained by Rayleigh--Ritz
minimization --- is used to compute ground and first-excited
eigenstates across a range of barrier depths. The design is
deliberately modest and interpretable in one dimension; it is not
intended as a contribution to variational methodology and does not
approach the deep architectures used for many-electron systems
\citep{carleo2017science,pfau2020ferminet,hermann2020paulinet}.
The trained wavefunctions are projected onto the harmonic-oscillator
basis, the Bargmann-polynomial zeros are extracted by numerical
root-finding, and the barrier parameter is swept continuously to
track zero migration through the tunneling transition.

The remainder of the paper is organized as follows.
Section~\ref{sec:theory} recalls the Bargmann representation and its
zero structure. Section~\ref{sec:methods} describes the variational
ansatz, training procedure, Bargmann projection, and barrier-sweep
protocol. Section~\ref{sec:results} reports numerical results.
Section~\ref{sec:discussion} discusses the physical interpretation of
imaginary-axis condensation, and Section~\ref{sec:conclusion}
summarizes.

\section{Theoretical background}
\label{sec:theory}

\subsection{The Bargmann representation}

Given a wavefunction $\psi(x) \in L^2(\mathbb{R})$ expanded in the
harmonic-oscillator basis
\begin{equation}
\begin{aligned}
   \psi(x) &= \sum_{n=0}^{\infty} c_n \, \phi_n(x), \\
   \phi_n(x) &= \frac{H_n(x)}{\sqrt{2^n n!\sqrt{\pi}}} \, e^{-x^2/2},
\end{aligned}
   \label{eq:ho_basis}
\end{equation}
where $H_n$ is the $n$-th Hermite polynomial, the Bargmann transform
of $\psi$ is the entire function
\begin{equation}
   \psi(z) = \sum_{n=0}^{\infty} \frac{c_n}{\sqrt{n!}} \, z^n,
   \qquad z \in \mathbb{C}.
   \label{eq:bargmann}
\end{equation}
The map is an isometric isomorphism between $L^2(\mathbb{R})$ and the
Bargmann--Fock space of entire functions square-integrable with respect
to the Gaussian measure $\pi^{-1} e^{-|z|^2}\,dz\,d\bar z$
\citep{bargmann1961fock}.

\subsection{Zeros and factorization}

By Hadamard's factorization theorem, an entire function $f(z)$ of
finite order is determined up to an exponential polynomial prefactor by
its zeros $\{z_k\}$ and its order. Truncating
equation~\eqref{eq:bargmann} to $n \leq N$ produces a polynomial with
exactly $N$ zeros in $\mathbb{C}$ (counted with multiplicity), some of
which approximate zeros of the full entire function while others are
spurious artifacts of the truncation. In practice, the zeros lying
well inside some fixed radius $R \ll \sqrt{N}$ are stable under
increase of $N$, and we restrict attention to these throughout.

Two symmetries are important. First, since $\psi(x) \in \mathbb{R}$ for
all physical eigenstates considered here, the coefficients $c_n$ are
real and the zero set is symmetric under complex conjugation
$z \mapsto \bar z$. Second, if $\psi(x)$ has a definite parity
($\psi(-x) = \pm \psi(x)$), then $c_n = 0$ for $n$ of the opposite
parity, which forces the Bargmann polynomial to be either even or odd
in $z$. An odd polynomial in $z$ has a zero at the origin and its
remaining zeros come in pairs $\{z, -z\}$; an even polynomial has its
zeros in such pairs. Combined with conjugation symmetry, parity
states therefore have zeros that either lie at the origin, on the
imaginary or real axis, or in groups of four $\{\pm z, \pm \bar z\}$.

\subsection{Nondimensionalization}
\label{sec:nondim}

Throughout this paper we adopt natural oscillator units
$\hbar = m = \omega = 1$. In these units, length is measured in
$x_0 \equiv \sqrt{\hbar/(m\omega)}$, energy in
$E_0 \equiv \hbar\omega$, and time in $\omega^{-1}$; all results below
are reported as dimensionless numbers to be multiplied by these
scales when interpreted for a physical system of given
$(\hbar, m, \omega)$.

Starting from the dimensional Schr\"odinger equation
$[-\hbar^2/(2m)\, \partial_x^2 + V(x)]\,\psi = E\,\psi$ and the
rescaling $x \to \tilde x = x/x_0$, the operator reduces to
$\tilde H = -\tfrac{1}{2}\,\partial_{\tilde x}^2 + \tilde V(\tilde x)$
with $\tilde V = V/E_0$ and $\tilde E = E/E_0$. The harmonic potential
becomes $\tilde V(\tilde x) = \tfrac{1}{2}\tilde x^2$, the anharmonic
becomes $\tilde V = \tfrac{1}{2}\tilde x^2 + \lambda \tilde x^4$ with
$\lambda$ dimensionless, and the double-well becomes
$\tilde V = \tfrac{1}{4}(\tilde x^2 - \tilde a^2)^2$ with barrier
parameter $\tilde a = a/x_0$ dimensionless. For compactness we drop
the tilde from now on: all quantities $(x, V, E, a, \varepsilon, z,
\sigma, \{c_k\}, \{\sigma_k\}, \{w_k\})$ below should be read as
dimensionless ratios to their natural units.

The tunneling splitting $\Delta$ and all energy errors are reported
in Hartree ($E_0 = \hbar\omega$); the Bargmann variable $z = (x +
ip)/\sqrt{2}$ carries no physical dimension in these units, and the
truncation order $N_\mathrm{max}$ is a pure integer counting
harmonic-oscillator Fock components.

\section{Methods}
\label{sec:methods}

All numerical work is performed on a one-dimensional grid
$x \in [-L, L]$ with $L = 8$ and $N_x = 1024$ equally spaced points.
Natural (Hartree) units are used throughout: $\hbar = m = 1$. We
consider three families of potentials:
\begin{align}
   \text{harmonic:}     &\quad V(x) = \tfrac{1}{2}\, x^2, \\
   \text{anharmonic:}   &\quad V(x) = \tfrac{1}{2}\, x^2 + \lambda\, x^4,
                            \quad \lambda > 0, \\
   \text{double-well:}  &\quad V(x) = \tfrac{1}{4}(x^2 - a^2)^2,
                            \quad a > 0.
\end{align}

\subsection{Finite-difference reference solver}
\label{sec:fd}

A three-point Dirichlet finite-difference discretization of the kinetic
operator, $T_{ij} = -\tfrac{1}{2}\,[D^{(2)}]_{ij}$, combined with the
diagonal potential matrix $V_{ij} = V(x_i)\,\delta_{ij}$, yields a
sparse symmetric Hamiltonian whose two lowest eigenpairs are obtained
by Lanczos iteration (\texttt{scipy.sparse.linalg.eigsh} from SciPy
\citep{virtanen2020scipy}, using NumPy arrays throughout
\citep{harris2020numpy}). Eigenvectors are normalized with the
discrete inner product $dx \cdot \sum_i$
to match the quadrature used downstream. This provides the reference
energies $\{E_0, E_1\}$ and wavefunctions against which the
variational ansatz is validated.

\subsection{Variational ansatz}
\label{sec:nn_ansatz}

The trial wavefunction is
\begin{equation}
   \psi_\theta(x) \,=\, f_\mathrm{sym}(x;\,\boldsymbol{\mu}) \cdot
   \left[\,1 + \varepsilon \cdot \tfrac{1}{2}
      \left( M_\theta(x) + p\, M_\theta(-x) \right)\right],
   \label{eq:ansatz}
\end{equation}
where $p \in \{+1, -1\}$ is the parity sector, $M_\theta$ is a
small feedforward correction network (4 hidden layers, 128 $\tanh$
units per layer), $\varepsilon$ is a learnable scalar, and
$f_\mathrm{sym}$ is a
physically motivated symbolic envelope multiplied by a polynomial
prefactor $1 + \sum_{k\geq 1} c_k u^k$, with $u = x^2$ for the
harmonic and anharmonic cases and $u = x^2 - a^2$ for the double-well.
For the harmonic and anharmonic cases, $f_\mathrm{sym}(x) = e^{-x^2/(2\sigma^2)}$
for even parity and $x \cdot e^{-x^2/(2\sigma^2)}$ for odd parity; for
the double-well, $f_\mathrm{sym}$ is a parity-projected sum of
three Gaussian components, each pair symmetrized around $\pm a$, with
trainable widths $\{\sigma_k\}$ and mixing weights $\{w_k\}$.

The parameters $\boldsymbol{\theta}$ (correction-term weights, $\varepsilon$,
$\sigma$, $\{c_k\}$, and for the double-well also $a$, $\{\sigma_k\}$,
$\{w_k\}$) are trained by Rayleigh--Ritz minimization of
\begin{equation}
   E[\psi_\theta] \,=\, \frac{\langle \psi_\theta | H | \psi_\theta \rangle}
                         {\langle \psi_\theta | \psi_\theta \rangle},
\end{equation}
evaluated on the same finite-difference grid and with the same discrete
Hamiltonian as Section~\ref{sec:fd}. The kinetic expectation value is
computed directly from the three-point stencil applied to
$\psi_\theta(x_i)$, so that the variational minimum is guaranteed to
coincide with the reference eigenvalue to machine precision if the
ansatz is expressive enough. First-excited states (odd parity) are
optimized in the same framework, with a Gram--Schmidt orthogonality
penalty $\alpha\,|\langle \psi_\theta | \psi_0 \rangle|^2$ added to the
loss, $\alpha = 50$.

Optimization proceeds in two stages. A four-phase Adam
\citep{kingma2015adam} schedule (learning rates
$3 \times 10^{-3} \to 1 \times 10^{-3} \to 3 \times 10^{-4} \to 5 \times 10^{-5}$,
10{,}000 total steps) is followed by an L-BFGS polish with strong-Wolfe line
search (4 outer iterations, 50 inner iterations each). For the
double-well first-excited state, where the optimizer is prone to a
local minimum roughly $6 \times 10^{-4}$ Ha above the ground truth, we
additionally use 4 random restarts and retain the best. All training
is implemented in PyTorch \citep{paszke2019pytorch}.

\subsection{Training disclosure}
\label{sec:training_disclosure}

All training is implemented in PyTorch 2.6
\citep{paszke2019pytorch} with 64-bit floating-point arithmetic
(\texttt{torch.float64}) throughout, so that energy and wavefunction
norms remain limited by the $h^2$ finite-difference discretization
error rather than accumulated floating-point noise. Optimization
proceeds in two stages: a four-phase Adam \citep{kingma2015adam}
schedule of $10{,}000$ full-batch gradient steps followed by an
L-BFGS polish with strong-Wolfe line search. All ablation tables
report mean $\pm$ std across 20 independent seeds per configuration,
generated by independent \texttt{torch.Generator} instances. Full
hyperparameter, initialization, reproducibility, and hardware details
are provided in Appendix~\ref{app:training_details}.

\subsection{Bargmann projection and zero extraction}
\label{sec:bargmann}

The harmonic-oscillator basis is evaluated on the grid via the
numerically stable three-term recurrence
\begin{equation}
   \phi_{n+1}(x) \,=\, \sqrt{\frac{2}{n+1}}\, x\, \phi_n(x)
   \,-\, \sqrt{\frac{n}{n+1}}\, \phi_{n-1}(x),
\end{equation}
starting from $\phi_0(x) = \pi^{-1/4} e^{-x^2/2}$ and
$\phi_1(x) = \sqrt{2}\, x\, \phi_0(x)$. Coefficients are obtained by
the discrete inner product $c_n = dx \sum_i \phi_n(x_i)\, \psi(x_i)$
for $n = 0, \ldots, N_\mathrm{max}$ with $N_\mathrm{max} = 30$, and the
Bargmann coefficients $a_n = c_n / \sqrt{n!}$ are computed in
log-space via $\sqrt{n!} = \exp(\tfrac{1}{2} \ln \Gamma(n+1))$ to avoid
overflow.

Before root-finding, a relative noise-floor threshold
$|a_n| < 10^{-4} \cdot \max_n |a_n|$ is applied: coefficients below
this threshold are set to zero and trailing zeros are trimmed. The
threshold is set above the numerical-projection noise ($\sim 10^{-5}$)
but well below the physical signal, and prevents spurious high-degree
roots from polluting the zero set. Zeros of the resulting polynomial
are then obtained by \texttt{numpy.polynomial.polynomial.polyroots}
and filtered by $|z| < R$ for a choice of $R \in \{3, 6, 9\}$.

\subsection{Barrier sweep protocol}
\label{sec:sweep}

To map the tunneling transition, the double-well barrier parameter is
swept over 20 values, $a \in [0.5, 2.3]$, uniformly spaced. At each
$a$, both eigenpairs $(E_0, \psi_0)$ and $(E_1, \psi_1)$ are computed
by the finite-difference solver of Section~\ref{sec:fd} -- which we
have shown to agree with the variational ansatz to within
$\sim 10^{-6}$ Ha -- and the zero set within $|z| < 6$ is extracted
as in Section~\ref{sec:bargmann}. The tunneling splitting
$\Delta(a) = E_1(a) - E_0(a)$ is tabulated in parallel.

\section{Results}
\label{sec:results}

\subsection{Reference spectra and ansatz validation}
\label{sec:results:energies}

Table~\ref{tab:energies} reports the ground- and first-excited-state
energies obtained by the finite-difference reference solver and by the
variational ansatz for the three test Hamiltonians. The anharmonic
value $\lambda = 0.1$ matches the canonical tabulation of
\citep{simon1970anharmonic}; the double-well value $a = 1.5$ is chosen to sit in
the mid-tunneling regime (splitting $\Delta \sim 0.26$ Ha). Agreement
with the reference is $\leq 10^{-5}$ Ha for all six energies, with most
values matching to $10^{-7}$ Ha. This confirms that the ansatz
expressivity is not the limiting factor in subsequent analysis.

\begin{table}[!htbp]
\centering
\begin{tabular}{lcccc}
\toprule
System & $E_0^\mathrm{FD}$ & $E_0^\mathrm{VA}$ &
         $E_1^\mathrm{FD}$ & $E_1^\mathrm{VA}$ \\
\midrule
harmonic     & 0.499992 & 0.499992 & 1.499962 & 1.499962 \\
anharmonic, $\lambda{=}0.1$   & 0.559135 & 0.559135 & 1.769438 & 1.769438 \\
double-well, $a{=}1.5$        & 0.801076 & 0.801076 & 1.062985 & 1.062987 \\
\bottomrule
\end{tabular}
\caption{Ground and first-excited state energies from the finite-difference
reference solver (FD) and the variational ansatz (VA) for
the three test Hamiltonians. Units: Hartree. FD values are exact at the
discretization level; VA values are the variational minimum after the
training schedule of Section~\ref{sec:nn_ansatz}.}
\label{tab:energies}
\end{table}

\begin{figure}[!htbp]
\centering
\includegraphics[width=0.7\textwidth]{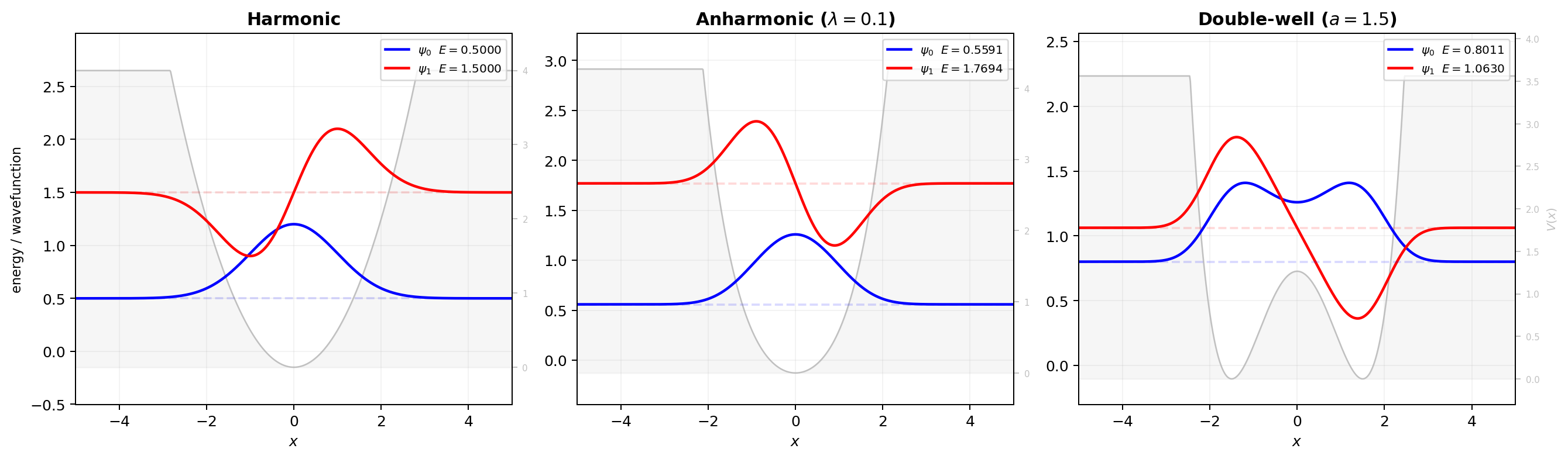}
\caption{\color{Gray} \textbf{Reference potentials and wavefunctions.}
Reference potentials and wavefunctions for the three test
Hamiltonians. Wavefunctions are rescaled for visibility and vertically
offset to their energies $E_0$, $E_1$. Horizontal dashed lines mark
the energy levels.}
\label{fig:phase0}
\end{figure}

Figure~\ref{fig:phase0} shows the reference potentials with the two
lowest eigenstates. The harmonic states are the standard Gaussian $e^{-x^2/2}$ and its
first-excited partner $x\,e^{-x^2/2}$; the anharmonic states are visually similar but
slightly tighter; the double-well ground state is the symmetric sum
and the first-excited state the antisymmetric difference of the two
lobes, with the small energy splitting $\Delta \approx 0.26$ Ha
characteristic of the mid-tunneling regime.

\begin{figure}[!htbp]
\centering
\includegraphics[width=0.7\textwidth]{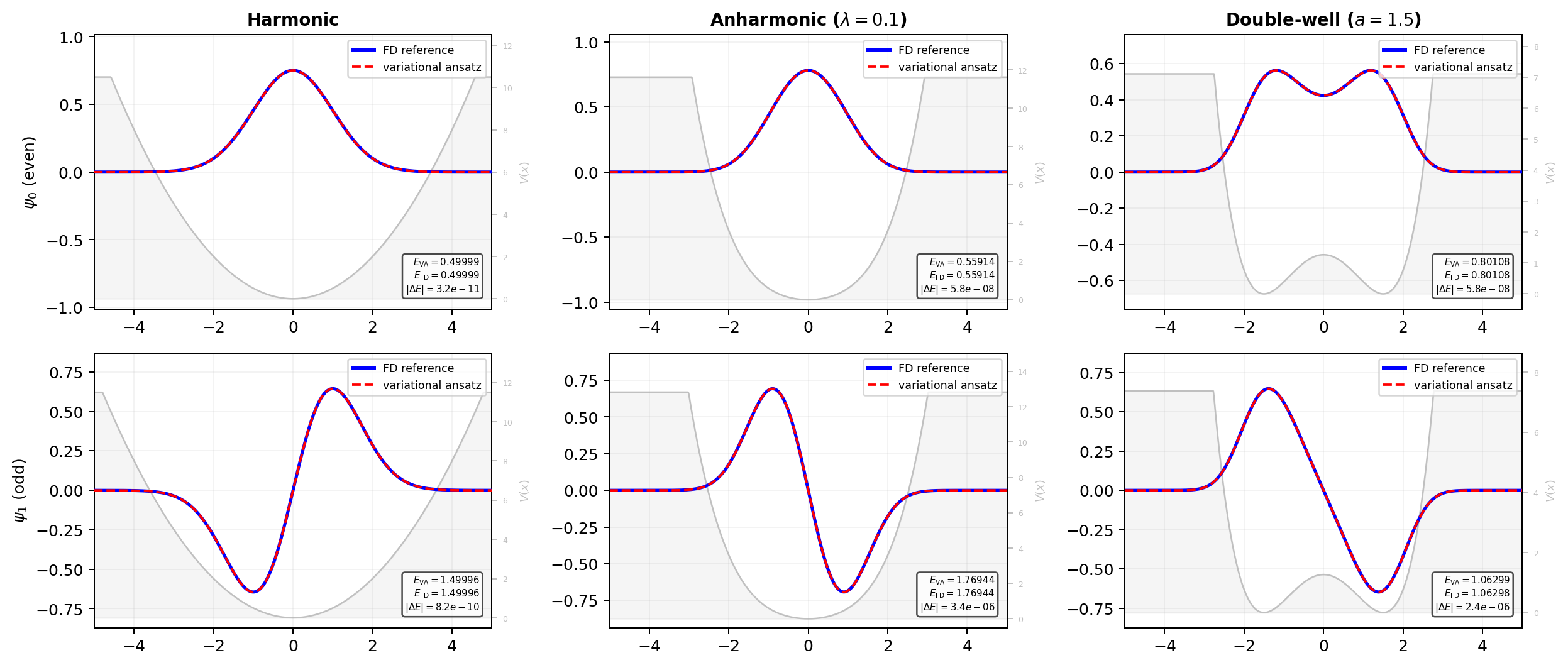}
\caption{\color{Gray} \textbf{Variational wavefunctions.}
Variational wavefunctions (red dashed) versus
the finite-difference reference (blue solid), with the potential
$V(x)$ overlaid in grey. Rows: ground state ($\psi_0$) and first
excited state ($\psi_1$). Columns: harmonic, anharmonic, double-well.
Agreement is within line-thickness throughout.}
\label{fig:phase1}
\end{figure}

Figure~\ref{fig:phase1} overlays the variational wavefunctions
(dashed red) on the finite-difference reference (solid blue) across
the three families. Residuals are at the plotting-line width, i.e.
indistinguishable by eye.

\FloatBarrier
\subsection{Bargmann zero galleries across the three families}
\label{sec:results:galleries}

\begin{figure}[!htbp]
\centering
\includegraphics[width=0.7\textwidth]{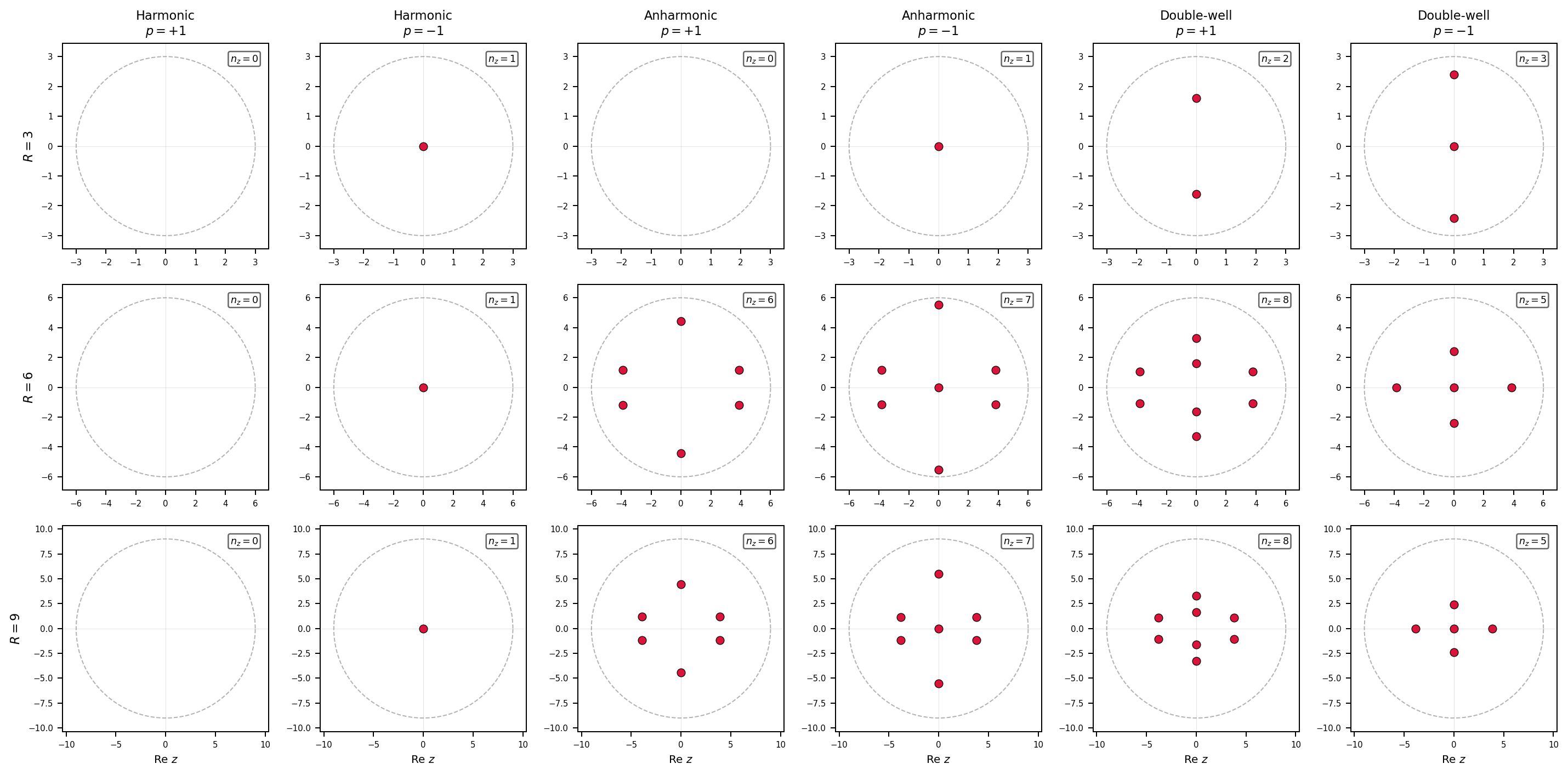}
\caption{\color{Gray} \textbf{Bargmann-polynomial zero galleries.} 
Bargmann-polynomial zero galleries for the variational
wavefunctions. Columns: three systems $\times$ two parities. Rows:
viewing radii $R = 3, 6, 9$. The harmonic ground and first-excited
states produce no zeros inside $R = 9$ -- they are essentially pure
Fock states. The anharmonic states produce a small number of zeros at
intermediate $|z|$. The double-well states (last two columns) produce
zeros localized \emph{on the imaginary axis}, including a zero at the
origin for the odd-parity first-excited state.}
\label{fig:phase2}
\end{figure}

With the trained wavefunctions validated, we project them onto the
harmonic-oscillator basis and extract the zeros of the truncated
Bargmann polynomial. Figure~\ref{fig:phase2} shows the zero galleries
for all six (system, parity) combinations at three viewing radii
$R \in \{3, 6, 9\}$.

Three qualitative observations are immediate. First, the harmonic
eigenstates are pure Fock states up to numerical floor, with $c_0 = 1$
for the ground state and $c_1 = 1$ for the first-excited state, and
consequently produce no Bargmann zeros in the viewing region (a single
zero at the origin for the odd state). Second, the anharmonic
eigenstates have a more distributed Fock spectrum but still produce
only a handful of low-magnitude zeros, with a small imaginary-axis pair
appearing only at large $|z|$ ($\approx \pm 4.4i$ and $\approx \pm 5.5i$
for the even and odd parities respectively) alongside an off-axis
four-cluster at $|\mathrm{Re}\, z| \approx 3.9$.
Third -- and this is the central physical observation of the paper --
the double-well eigenstates produce zeros that \emph{lie on the
imaginary axis} (within numerical precision), both for the ground
state ($p = +1$, a pair of imaginary zeros at $z \approx \pm 1.6i$)
and for the first-excited state ($p = -1$, a zero at the origin plus
an imaginary pair at $z \approx \pm 2.4i$).

The parity-conjugation symmetry analysis of Section~\ref{sec:theory}
predicts that double-well eigenstates, being real and of definite
parity, must have zeros either at the origin, on the real axis, on
the imaginary axis, or in groups of four. That they preferentially
occupy the imaginary axis rather than the real axis is a separate
physical statement, distinguishing tunneling-like from pure
anharmonic-like behavior.

\FloatBarrier
\subsection{Zero migration across the tunneling transition}
\label{sec:results:sweep}

\begin{figure}[!htbp]
\centering
\includegraphics[width=0.55\textwidth]{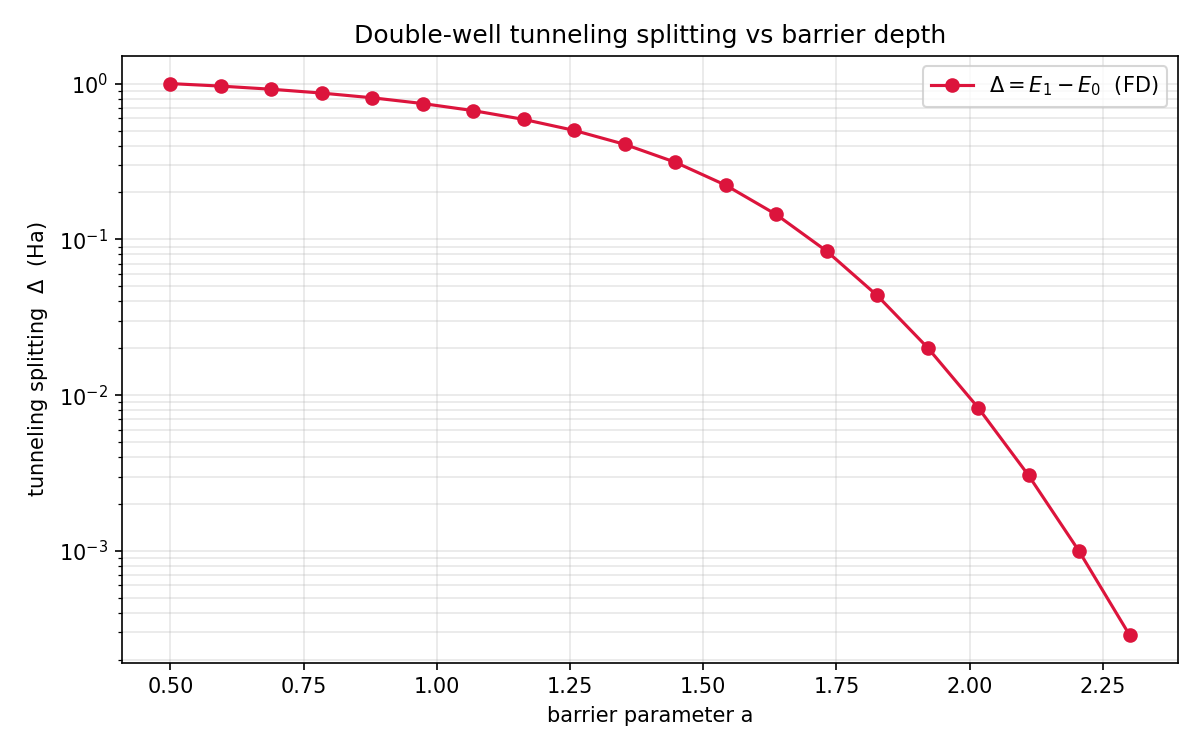}
\caption{\color{Gray} \textbf{Tunneling splitting.}
Tunneling splitting $\Delta = E_1 - E_0$ of the symmetric
double-well potential $V(x) = \tfrac{1}{4}(x^2 - a^2)^2$ as a function
of barrier parameter $a$. Logarithmic vertical scale; 20 uniformly
spaced values of $a$. The $3.5$-decade descent of $\Delta(a)$ over
$a \in [0.5, 2.3]$ is consistent with an instanton-action suppression
$\Delta \sim \exp(-S_0(a))$.}
\label{fig:splitting}
\end{figure}

We now sweep the barrier parameter $a$ continuously and track how
the zero set evolves.
Figure~\ref{fig:splitting} shows the tunneling splitting
$\Delta(a) = E_1 - E_0$ on a logarithmic vertical scale. The splitting
falls from $\Delta \approx 1.0$ Ha at $a = 0.5$ to
$\Delta \approx 3 \times 10^{-4}$ Ha at $a = 2.3$, a range of
$3.5$ decades. The curvature on the log scale is consistent with the
WKB/instanton prediction \citep{garg2000tunnel,zinnjustin2004multiinstanton},
based on the standard methodology of Coleman \citep{coleman1977fate},
that the splitting scales as $\Delta(a) \sim \exp(-S_0(a))$ with an
action $S_0$ that grows faster than linearly with $a$ as the barrier
deepens.

Figure~\ref{fig:traj} shows the imaginary parts $\mathrm{Im}\, z$ of
all Bargmann zeros within $|z| < 6$, plotted against $a$ and colored
by the absolute real part $|\mathrm{Re}\, z|$. Points near zero (dark
purple) are on the imaginary axis; greener and yellower points are
displaced from it. Several features are visible:

\begin{figure}[!htbp]
\centering
\includegraphics[width=0.62\textwidth]{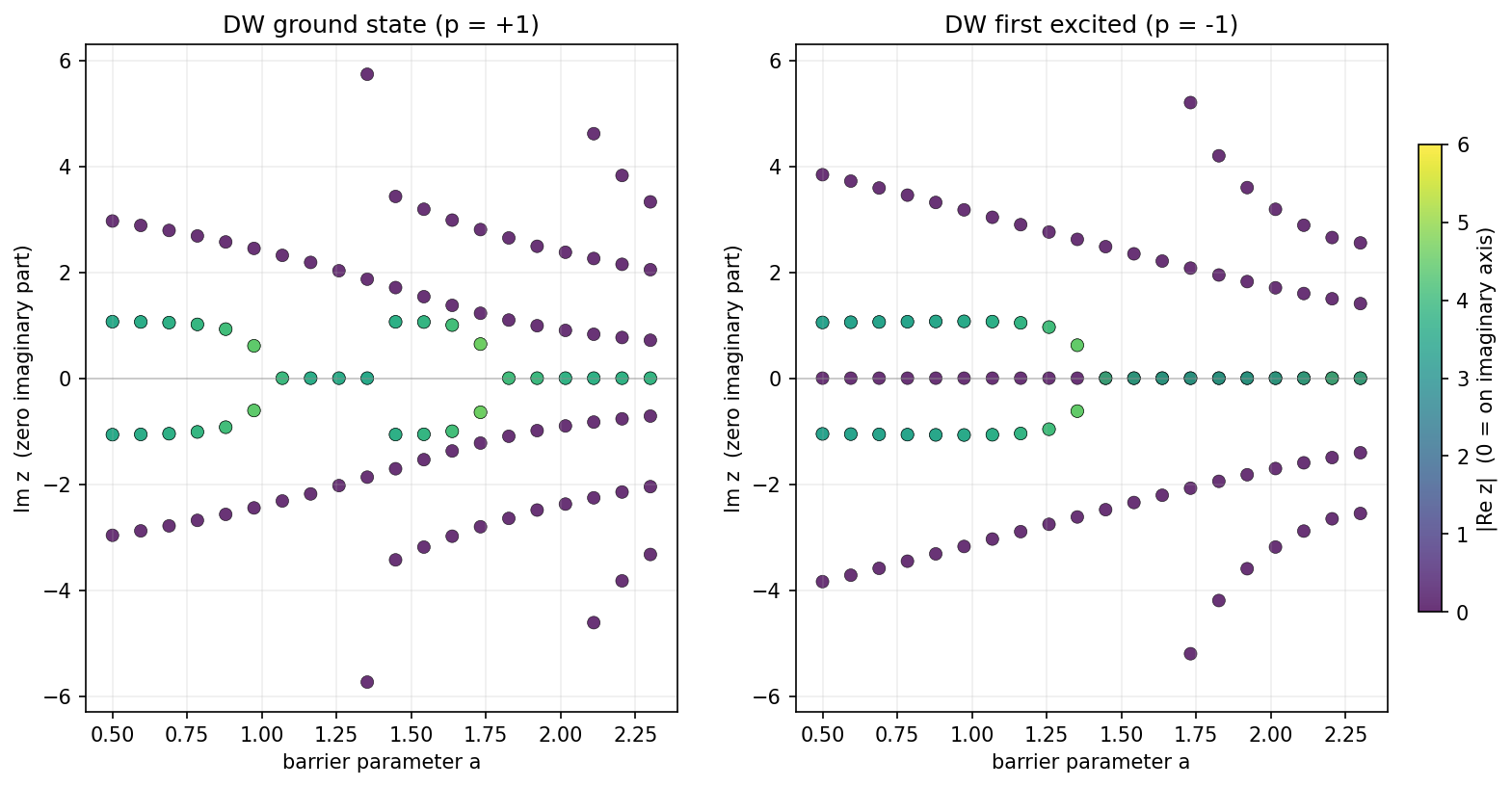}
\caption{\color{Gray} \textbf{Imaginary parts of Bargmann zeros.}
Imaginary parts of Bargmann zeros (within $|z| < 6$) of the
double-well ground state (left, $p = +1$) and first-excited state
(right, $p = -1$), as a function of barrier parameter $a$. Points are
colored by $|\mathrm{Re}\, z|$; points with $|\mathrm{Re}\, z| \approx 0$
(dark purple) lie on the imaginary axis. The transition from off-axis
to on-axis localization proceeds through the interval
$a \in [1.0, 1.5]$.}
\label{fig:traj}
\end{figure}

Three stages of the transition are visible.
At small $a$ ($a \lesssim 1.0$), the zero set comprises four off-axis
points (bright green/yellow) together with two zeros already on the
imaginary axis (dark purple), arranged in a near-hexagonal pattern
consistent with a near-harmonic state carrying small Fock components
at high $n$. Around $a \approx 1.0$--$1.3$ some off-axis zeros
collapse onto the imaginary axis while others migrate outward beyond
$R = 6$ and leave the tracked set. At $a \gtrsim 1.5$, nearly all
zeros within the viewing radius lie on the imaginary axis, and their
number grows with $a$ as previously off-axis zeros are drawn
onto the axis.

The qualitative correspondence between Fig.~\ref{fig:splitting} and
Fig.~\ref{fig:traj} is that the exponential collapse of $\Delta(a)$
and the imaginary-axis condensation of the Bargmann zeros are two
views of the same underlying change in the wavefunction's Fock-space
structure.

\begin{figure}[!htbp]
\centering
\includegraphics[width=0.62\textwidth]{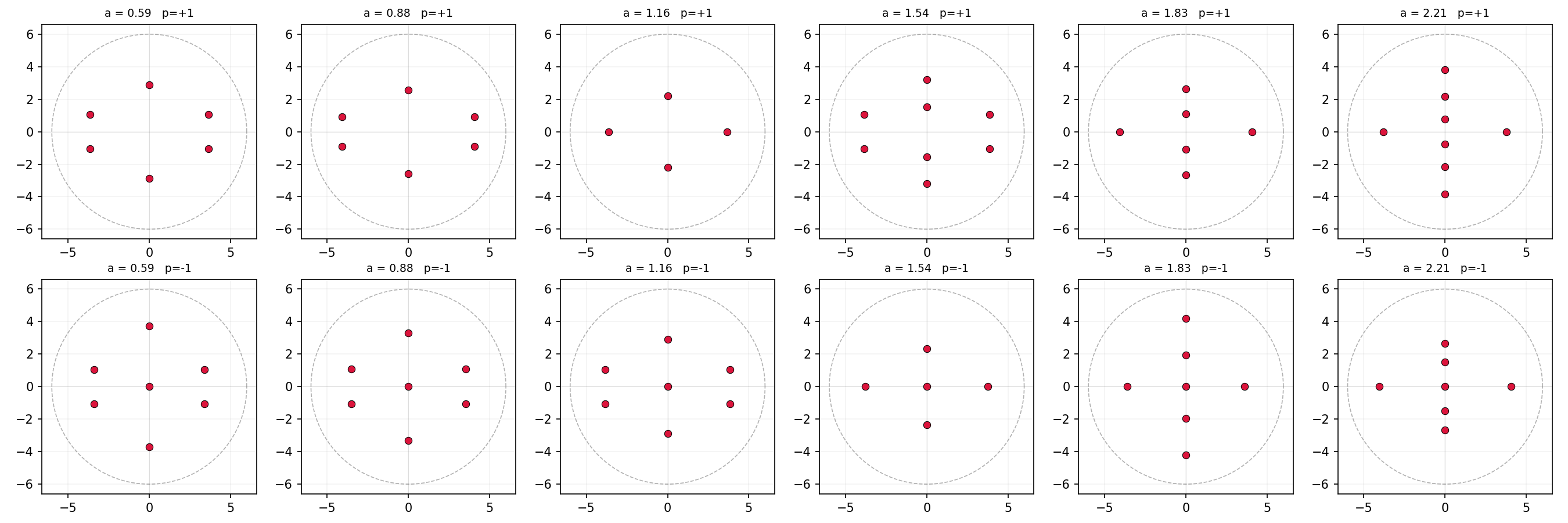}
\caption{\color{Gray} \textbf{Snapshots of the Bargmann zero set.}
Snapshots of the Bargmann zero set in the complex plane at
six selected values of the barrier parameter $a$. Top row: ground
state ($p = +1$); bottom row: first excited state ($p = -1$). Dashed
circle indicates the viewing radius $R = 6$. The imaginary-axis
condensation is visible in the progression from $a = 0.59$ (rotationally
symmetric) to $a = 2.21$ (zeros stacked vertically on the imaginary
axis).}
\label{fig:panels}
\end{figure}

Figure~\ref{fig:panels} shows six snapshot panels of the zero set in
the complex plane at selected values of $a$. The near-harmonic regime
at $a = 0.59$ exhibits a roughly hexagonal pattern with four off-axis
zeros and two already on the imaginary axis; by
$a = 2.21$ the zeros are stacked on the imaginary axis with only
two zeros surviving at large $|\mathrm{Re}\, z| \approx 3.8$.

\subsection{Ablation studies}
\label{sec:results:ablation}

To establish the robustness of the numerical results reported above, we
perform systematic ablation studies varying four key methodological
choices: the spatial grid resolution, the correction-network capacity, the
perturbation strength~$\varepsilon$, and the Bargmann truncation order.

\subsubsection{Grid-resolution ablation}
\label{sec:results:ablation:grid}

Table~\ref{tab:ablation_grid} reports the ground- and first-excited-state
energies across four grid resolutions
$N \in \{512, 1024, 2048, 4096\}$ for all six test potentials.
The energy drift between $N=1024$ and $N=4096$ is $\leq 5 \times 10^{-5}$~Ha
across all potentials, consistent with the expected $h^2$ convergence of
the three-point finite-difference stencil. This justifies $N=1024$ as the
default grid resolution throughout the paper.

\begin{table}[!htbp]
\centering
\small
\setlength{\tabcolsep}{5pt}
\renewcommand{\arraystretch}{0.82}
\begin{tabular}{llcccc}
\toprule
$N$ & Potential & $E_0$ & $E_1$ & $\Delta$ & $t$ (s) \\
\midrule
512  & harmonic         & 0.499969 & 1.499847 & 0.999877 & 0.05 \\
512  & anharmonic 0.1   & 0.559103 & 1.769244 & 1.210142 & 0.04 \\
512  & anharmonic 0.5   & 0.696097 & 2.323884 & 1.627787 & 0.04 \\
512  & DW $a{=}1.0$     & 0.397210 & 1.122084 & 0.724874 & 0.05 \\
512  & DW $a{=}1.5$     & 0.801030 & 1.062885 & 0.261855 & 0.06 \\
512  & DW $a{=}2.0$     & 1.338342 & 1.348060 & 0.009718 & 0.04 \\
\midrule
1024 & harmonic         & 0.499992 & 1.499962 & 0.999969 & 0.14 \\
1024 & anharmonic 0.1   & 0.559135 & 1.769438 & 1.210303 & 0.12 \\
1024 & anharmonic 0.5   & 0.696156 & 2.324276 & 1.628120 & 0.11 \\
1024 & DW $a{=}1.0$     & 0.397229 & 1.122217 & 0.724988 & 0.16 \\
1024 & DW $a{=}1.5$     & 0.801076 & 1.062985 & 0.261909 & 0.19 \\
1024 & DW $a{=}2.0$     & 1.338500 & 1.348216 & 0.009717 & 0.13 \\
\midrule
2048 & harmonic         & 0.499998 & 1.499990 & 0.999992 & 0.72 \\
2048 & anharmonic 0.1   & 0.559144 & 1.769487 & 1.210343 & 0.57 \\
2048 & anharmonic 0.5   & 0.696171 & 2.324374 & 1.628203 & 0.43 \\
2048 & DW $a{=}1.0$     & 0.397234 & 1.122250 & 0.725017 & 0.74 \\
2048 & DW $a{=}1.5$     & 0.801087 & 1.063010 & 0.261923 & 0.92 \\
2048 & DW $a{=}2.0$     & 1.338539 & 1.348255 & 0.009716 & 0.59 \\
\midrule
4096 & harmonic         & 0.500000 & 1.499998 & 0.999998 & 4.61 \\
4096 & anharmonic 0.1   & 0.559146 & 1.769499 & 1.210353 & 3.58 \\
4096 & anharmonic 0.5   & 0.696175 & 2.324398 & 1.628224 & 2.63 \\
4096 & DW $a{=}1.0$     & 0.397235 & 1.122258 & 0.725024 & 4.40 \\
4096 & DW $a{=}1.5$     & 0.801090 & 1.063016 & 0.261926 & 5.62 \\
4096 & DW $a{=}2.0$     & 1.338549 & 1.348265 & 0.009716 & 3.73 \\
\bottomrule
\end{tabular}
\caption{Grid-resolution ablation. Ground-state energy $E_0$,
first-excited-state energy $E_1$, and tunneling splitting
$\Delta = E_1 - E_0$ for six potentials at four grid sizes.
Units: Hartree. Wall-clock time per solve in seconds.}
\label{tab:ablation_grid}
\end{table}

\subsubsection{Correction-network capacity ablation}
\label{sec:results:ablation:mlp}

To verify that the paper's conclusions are insensitive to the choice
of correction-network architecture, we sweep over a $3 \times 4$ grid
of configurations (depths $\{2, 3, 4\}$, widths $\{16, 32, 64,
128\}$) on the double-well potential at $a = 1.5$, with 20 random
seeds per configuration. Even the smallest architecture
($2 \times 16$, 331 parameters) achieves a mean energy error of
$8.97 \times 10^{-7}$~Ha, more than two orders of magnitude below
the $10^{-5}$~Ha threshold of the abstract; the largest
($4 \times 128$, 49{,}931 parameters) reaches
$1.9 \times 10^{-8}$~Ha and saturates near the $N=1024$
grid-resolution floor. The paper's qualitative conclusions are
therefore architecturally insensitive; the $4 \times 128$
configuration is used as the default elsewhere in the paper. Full
results are provided in Appendix~\ref{app:arch_ablation},
Table~\ref{tab:ablation_mlp}.

\subsubsection{Perturbation strength ablation}
\label{sec:results:ablation:epsilon}

A central design choice of the variational ansatz \eqref{eq:ansatz}
is that the correction term enters multiplicatively as a small
perturbation of the symbolic envelope, scaled by a dimensionless
parameter $\varepsilon$. We sweep $\varepsilon$ across fifteen
log-spaced values from $0.005$ to $0.7$, with 20 seeds each
(holding $\varepsilon$ fixed at each swept value during training,
rather than letting it refine as in the main runs), and
extract the leading Bargmann zero $|z_0|$ at each value. The energy
error decreases smoothly from $\sim 3 \times 10^{-7}$~Ha at
$\varepsilon = 0.005$ to below $10^{-8}$~Ha at $\varepsilon \geq
0.1$, while the leading-zero position is pinned at
$|z_0| = 1.6143$ across all fifteen values to four decimal places,
with seed-to-seed standard deviation $\sim 10^{-4}$ and no detectable
systematic drift with $\varepsilon$. The correction term thus
improves the variational energy without deforming the Bargmann-zero
topology; $\varepsilon$ is genuinely perturbative with respect to
the paper's central diagnostic, and the results of
Section~\ref{sec:results:sweep} depend only on the symbolic
envelope's correct parity and locality structure. Full results are
provided in Appendix~\ref{app:epsilon_ablation},
Table~\ref{tab:ablation_epsilon}.

\subsubsection{Bargmann truncation ablation}
\label{sec:results:ablation:truncation}

The Bargmann representation \eqref{eq:bargmann} is formally an
infinite series in $z$; for a numerical implementation we truncate at
a finite Fock order $N_\mathrm{max}$. We use $N_\mathrm{max} = 30$
throughout the paper (Section~\ref{sec:bargmann}); to verify that
this choice is conservative and does not impose a systematic error
on the zero-set analysis, we sweep $N_\mathrm{max} \in \{20, 30, 40,
60, 80, 100, 120, 160, 200\}$ and compare the resulting zero sets
against a high-$N_\mathrm{max}$ reference at $N_\mathrm{max} = 200$.
The sweep is run over one wavefunction per $\varepsilon$ value
(seed 0, 15 wavefunctions total) from the $\varepsilon$ ablation of
Section~\ref{sec:results:ablation:epsilon}, and drift is measured by
nearest-neighbour matching of the eight leading zeros
($|z|$-ordered) to their reference positions at $N_\mathrm{max} =
200$.

The result is essentially trivial: for all $N_\mathrm{max} \geq 30$
the zero positions agree with the reference to machine precision
(drift identically zero to within the floating-point root-finding
algorithm's own tolerance), and at $N_\mathrm{max} = 20$ the residual
drift is $\sim 10^{-14}$, at numerical-noise level
(Table~\ref{tab:ablation_truncation}). The low-frequency Fock
components of the double-well ground state dominate its Bargmann
representation so strongly that keeping even only the first 20
components is sufficient to localize every zero in the viewing
region $|z| < 6$. This confirms that the paper's
$N_\mathrm{max} = 30$ default is very safe; even
$N_\mathrm{max} = 20$ would be defensible in applications where the
projection cost matters.

\begin{table}[!htbp]
\centering
\begin{tabular}{ccc}
\toprule
$N_\mathrm{max}$ & Mean drift vs.\ $N_\mathrm{max}{=}200$
                 & Max drift \\
\midrule
20  & $7.62 \times 10^{-15}$ & $1.51 \times 10^{-14}$ \\
30  & $0$ (exact)            & $0$ \\
40  & $0$ (exact)            & $0$ \\
60  & $0$ (exact)            & $0$ \\
80  & $0$ (exact)            & $0$ \\
100 & $0$ (exact)            & $0$ \\
120 & $0$ (exact)            & $0$ \\
160 & $0$ (exact)            & $0$ \\
200 & $0$ (reference)        & $0$ \\
\bottomrule
\end{tabular}
\caption{Bargmann truncation ablation. Zero-set drift
(mean over 15 $\varepsilon$ values at seed 0) between the leading
eight zeros at $N_\mathrm{max}$ and at $N_\mathrm{max} = 200$, using
nearest-neighbour matching. Drift at $N_\mathrm{max} \geq 30$ is
zero to within the root-finder's own tolerance; the residual at
$N_\mathrm{max} = 20$ is at double-precision numerical-noise level.}
\label{tab:ablation_truncation}
\end{table}

\subsubsection{Wavefunction-level $L_2$ validation}
\label{sec:results:ablation:l2}

The energy-level agreement reported in
Section~\ref{sec:results:energies} is a necessary but not sufficient
condition for trusting the Bargmann-zero analysis, because the zero
set depends on the full wavefunction shape and not on the energy
alone. To provide a direct measure of wavefunction-level fidelity,
we compute the pointwise error
$\psi_\theta - \psi_\mathrm{FD}$ on the training grid and report its
$L_2$ and $L_\infty$ norms across all seven reference systems
(harmonic, anharmonic with $\lambda = 0.1$ and $\lambda = 0.5$, and
double-well with $a \in \{1.0, 1.5, 2.0, 2.5\}$, the last point
extending slightly beyond the barrier sweep of
Section~\ref{sec:sweep} to probe the deep-tunneling regime), each
with 20 independent seeds. Both wavefunctions are normalized with the
discrete $dx\,\sum$ quadrature, and the sign of $\psi_\theta$ is
fixed to match $\psi_\mathrm{FD}$ before the subtraction.

Table~\ref{tab:l2_validation} summarizes the results. The mean
energy error is $\lesssim 6 \times 10^{-8}$~Ha across every
potential, the mean $L_2$ error is $\sim 1$--$3 \times 10^{-5}$,
and the worst-case $L_2$ error over any single seed is
$3.8 \times 10^{-5}$ (double-well $a = 1.0$). These errors lie
about an order of magnitude below the $10^{-4}$ threshold
commonly used as a variational-accuracy benchmark, and the $L_2$
floor is approximately constant across all seven systems, indicating
that the limiting factor is the $N = 1024$ finite-difference
discretization rather than any system-specific loss of ansatz
expressivity. Together with the energy-level and Bargmann-zero
agreement reported elsewhere, this confirms wavefunction-level
fidelity at the discretization floor.

\begin{table}[!htbp]
\centering
\begin{tabular}{lcccc}
\toprule
System & Mean $|\Delta E|$ (Ha)
       & Mean $\|\Delta\psi\|_2$
       & Max $\|\Delta\psi\|_2$
       & Mean $\|\Delta\psi\|_\infty$ \\
\midrule
harmonic                    & $4.75 \times 10^{-9}$ & $1.26 \times 10^{-5}$ & $2.77 \times 10^{-5}$ & $9.43 \times 10^{-6}$ \\
anharmonic, $\lambda{=}0.1$ & $2.59 \times 10^{-8}$ & $2.43 \times 10^{-5}$ & $3.26 \times 10^{-5}$ & $1.70 \times 10^{-5}$ \\
anharmonic, $\lambda{=}0.5$ & $6.08 \times 10^{-8}$ & $2.55 \times 10^{-5}$ & $3.52 \times 10^{-5}$ & $2.14 \times 10^{-5}$ \\
DW, $a{=}1.0$               & $3.62 \times 10^{-8}$ & $1.90 \times 10^{-5}$ & $3.84 \times 10^{-5}$ & $1.57 \times 10^{-5}$ \\
DW, $a{=}1.5$               & $1.25 \times 10^{-8}$ & $1.00 \times 10^{-5}$ & $2.49 \times 10^{-5}$ & $8.93 \times 10^{-6}$ \\
DW, $a{=}2.0$               & $1.26 \times 10^{-8}$ & $1.06 \times 10^{-5}$ & $1.66 \times 10^{-5}$ & $1.01 \times 10^{-5}$ \\
DW, $a{=}2.5$               & $1.86 \times 10^{-8}$ & $1.20 \times 10^{-5}$ & $2.05 \times 10^{-5}$ & $1.03 \times 10^{-5}$ \\
\bottomrule
\end{tabular}
\caption{Wavefunction-level $L_2$ and $L_\infty$ validation across
the seven reference systems, 20 seeds per system.
All errors are for the ground state ($p = +1$) on the $N=1024$ grid
after sign alignment and normalization. Reference: the corresponding
finite-difference eigenvector $\psi_\mathrm{FD}$ at the same grid.}
\label{tab:l2_validation}
\end{table}

\section{Discussion}
\label{sec:discussion}

\subsection{Physical origin of imaginary-axis condensation}

The imaginary-axis localization of Bargmann zeros for the double-well
eigenstates can be understood at the level of the Fock-coefficient
spectrum. Under the parity-projection condition $c_n = 0$ for odd $n$
(even-parity ground state) or even $n$ (odd-parity first-excited
state), the Bargmann polynomial takes the form
\begin{equation}
   \psi(z) \,=\, \sum_{k} \frac{c_{2k}}{\sqrt{(2k)!}}\, z^{2k}
   \qquad \text{or} \qquad
   z \sum_{k} \frac{c_{2k+1}}{\sqrt{(2k+1)!}}\, z^{2k},
\end{equation}
i.e. a polynomial in $z^2$ (or $z$ times a polynomial in $z^2$). The
transformation $w = z^2$ therefore contracts the $z$-plane zero-pair
$\{\zeta, -\zeta\}$ to a single $w$-plane zero $w_0 = \zeta^2$. Real
$w_0 < 0$ (i.e.\ zeros in the left half $w$-plane) correspond to pure
imaginary $\zeta$; real $w_0 > 0$ to pure real $\zeta$; complex $w_0$
to a four-fold cluster $\{\pm \zeta, \pm\bar \zeta\}$. The observed
condensation of $z$-zeros onto the imaginary axis is thus equivalent
to a condensation of $w$-zeros onto the negative real semi-axis. This
latter condition is, at least heuristically, a signature of a specific
sign structure of the Fock-coefficient sequence $\{c_{2k}\}$ that
arises precisely when the wavefunction is bimodally localized in
position space (as is the case in the deep-tunneling regime).

Equivalently, the imaginary-axis zeros may be interpreted as
Husimi-phase-space nodes of the coherent-state transform
$Q(z) = \pi^{-1}|\psi(z)|^2\, e^{-|z|^2}$, with $Q$ as originally defined by
Husimi \citep{husimi1940formal} and subsequently developed within the
Glauber--Sudarshan coherent-state framework
\citep{glauber1963coherent,klauder1985coherent}.
This connects the present observation to the programme
\citep{leboeuf1990zeros,cerf2025zeros} in which the zero set of the
holomorphic (or Husimi) representation is treated as the primary object
of study rather than the wavefunction itself. Leboeuf \& Voros
\citep{leboeuf1990zeros} demonstrated that the zero set of the Husimi
function is chaos-revealing for classically non-integrable systems, and
Cerf \emph{et al.} \citep{cerf2025zeros} have sharpened the
connection via a Hudson-type theorem linking zeros to non-Gaussianity.
We do not pursue that direction further here, but note that it provides
an independent computational route to the same diagnostic, useful when
the Bargmann polynomial is too high-degree for direct root-finding to
remain numerically stable.

\subsection{Limitations}

The present analysis is restricted to one-dimensional systems,
parity-symmetric states, and the two lowest eigenvalues. Extension to
higher excited states presents no methodological obstacle --- the
Bargmann polynomial becomes higher-degree and zero-finding more
expensive, but the procedure is unchanged --- though the parity-zero
symmetry that simplifies the present analysis will be broken for
asymmetric or multi-minima potentials. Extension to two-dimensional
systems requires a two-mode Bargmann representation whose zero set is
a variety in $\mathbb{C}^2$ rather than a discrete point set; this is
left to future work. Finally, the variational ansatz is modest by the
standards of contemporary variational wavefunctions
\citep{carleo2017science,pfau2020ferminet,hermann2020paulinet};
it was chosen for interpretability in one dimension rather than as a
contribution to variational Monte Carlo.

\section{Conclusion}
\label{sec:conclusion}

We have shown that the complex zeros of the Bargmann-represented
wavefunction carry a recognizable signature of the tunneling
regime in the one-dimensional symmetric double well. A variational
ansatz reproducing finite-difference reference energies to
$\sim 10^{-5}$\,Ha is used to project ground and first-excited
eigenstates onto the harmonic-oscillator basis; the zeros of the
truncated Bargmann polynomial are then extracted across a continuous
sweep of the barrier parameter. The zeros migrate onto the imaginary
axis as the barrier deepens, concurrent with the $3.5$-decade
exponential collapse of the tunneling splitting. This imaginary-axis
condensation is traced to a sign-alternation structure in the
Fock-coefficient spectrum characteristic of bimodally localized
wavefunctions, and is connected to the Husimi-function zero set
through the coherent-state representation.

The result extends the random-polynomial Bargmann-zero framework
\citep{janowicz2022bargmann} to physically realized eigenstates,
identifying the double-well tunneling transition as a phenomenon
directly readable from the zero set. Generalization to asymmetric
double wells, to excited states with $n > 1$, and to multi-mode
systems where the Bargmann zero set becomes a higher-dimensional
algebraic variety are natural directions for subsequent work.

\section*{Acknowledgments}
The authors declare no competing interests.

During the preparation of this work, generative AI tools were used
in two distinct capacities. First, AI-assisted code generation
(GitHub Copilot and ChatGPT) was used for initial prototyping of
the variational training loop and the Bargmann-projection routines;
all final implementation was written, inspected, and verified by the
authors. Second, a large language model (ChatGPT) was used to assist
with prose editing and language refinement at the manuscript stage.
The authors reviewed and took responsibility for all content thus
produced. No AI tool is listed as an author; all authors accept full
responsibility for the scientific content of this work, consistent
with arXiv policy.

\section*{Funding}
No external funding was received for this work.

\section*{Author Contributions}
\textbf{Tughanbulut Kurtulush}: Conceptualization, Methodology, Software, Investigation, Formal analysis, 
Visualization, Writing -- original draft.
\textbf{Maciej Janowicz}: Conceptualization, Methodology, Supervision, 
Writing -- review \& editing.

\section*{Data Availability}
All code, trained model weights, and generated datasets supporting
the findings of this study are openly available at
\url{https://github.com/TuanBulut/bargmann-zeros-tunneling} and will be
archived with a Zenodo DOI upon acceptance. The
repository includes the finite-difference reference solver, the
variational training scripts, the Bargmann-projection and
zero-extraction utilities, the barrier-sweep driver, and the raw
zero-maps used to produce all figures and tables in this manuscript.


\appendix

\section{Training hyperparameters and hardware}
\label{app:training_details}

This appendix gives the full hyperparameter, initialization,
reproducibility, and hardware details summarized in
Section~\ref{sec:training_disclosure}.

\textit{Optimizer schedule.} Training proceeds in two stages. A
four-phase Adam \citep{kingma2015adam} schedule with default momenta
$\beta = (0.9, 0.999)$ and learning rates $3\times 10^{-3} \to
1\times 10^{-3} \to 3\times 10^{-4} \to 5\times 10^{-5}$ is applied
for a total of $10{,}000$ full-batch gradient steps (the grid
$x_i$ is the entire ``batch''); the learning rate is stepped down
after $30\%$, $60\%$, and $85\%$ of the schedule respectively. An
L-BFGS polish with strong-Wolfe line search then runs for up to
$4$ outer iterations of $50$ inner iterations each, with gradient
and parameter-change tolerances $10^{-12}$ and $10^{-14}$
respectively, and a Hessian history of size $50$. The best energy
encountered at any point during the schedule is retained;
backtracking to earlier checkpoints is not needed in practice
because the L-BFGS polish is always monotonically improving.

\textit{Initialization.} Correction-network weights are drawn
independently from $\mathcal{N}(0, 0.3^2)$ per layer; biases are
initialized to zero. Envelope width parameters $\{\sigma_k\}$ are
initialized logarithmically spaced in $[0.35, 1.0]$;
polynomial prefactor coefficients $\{c_k\}$ are drawn from
$\mathcal{N}(0, 0.01^2)$; Gaussian mixture weights $\{w_k\}$ are
initialized uniformly; the double-well barrier parameter inside the
ansatz is initialized at its target value (e.g.\ $a = 1.5$ for the
mid-tunneling baseline) and allowed to refine during training.

\textit{Reproducibility.} Each training run is seeded via an
independent \texttt{torch.Generator} instance, so that per-seed
initial conditions, and therefore per-seed final energies and
Bargmann zero positions, are deterministic. All ablation tables
report mean $\pm$ std across 20 independent seeds per configuration.

\textit{Hardware and wall-clock.} Energy-ablation runs reported in
Section~\ref{sec:results:ablation} were carried out on a single
NVIDIA GeForce RTX 4060 Laptop GPU (8\,GB VRAM) using FP64
throughout. Independent seeds (and, where applicable, independent
$\varepsilon$ or architecture configurations) were batched as a
leading tensor dimension, so that a single forward/backward pass
updates all configurations simultaneously; the L-BFGS polish runs
serially per-seed to preserve the per-seed curvature estimate. The
$15\varepsilon \times 20$-seed sweep of
Section~\ref{sec:results:ablation:epsilon} completed in
approximately $3\,\mathrm{h}$ (Adam) $+\, 16\,\mathrm{min}$
(L-BFGS) of wall-clock time on this hardware. The $L_2$ validation
across seven potentials completed in approximately
$70\,\mathrm{min}$.

\section{Correction-network capacity ablation: full results}
\label{app:arch_ablation}

Table~\ref{tab:ablation_mlp} reports the full $3\times 4$ grid of
correction-network configurations (depths $\{2, 3, 4\}$, widths
$\{16, 32, 64, 128\}$), trained on the double-well potential at
$a = 1.5$ (ground state, $p = +1$) with 20 independent random seeds
per configuration (240 runs total).

\begin{table}[!htbp]
\centering
\begin{tabular}{ccrcc}
\toprule
Depth & Width & Params & Mean $|\Delta E|$ (Ha) & Std (Ha) \\
\midrule
2 & 16  & 331      & $8.97 \times 10^{-7}$ & $2.10 \times 10^{-7}$ \\
2 & 32  & 1{,}163  & $6.24 \times 10^{-7}$ & $1.29 \times 10^{-7}$ \\
2 & 64  & 4{,}363  & $4.47 \times 10^{-7}$ & $2.26 \times 10^{-7}$ \\
2 & 128 & 16{,}907 & $2.07 \times 10^{-7}$ & $1.82 \times 10^{-7}$ \\
\midrule
3 & 16  & 603      & $6.81 \times 10^{-7}$ & $1.36 \times 10^{-7}$ \\
3 & 32  & 2{,}219  & $5.72 \times 10^{-7}$ & $1.56 \times 10^{-7}$ \\
3 & 64  & 8{,}523  & $1.95 \times 10^{-7}$ & $1.66 \times 10^{-7}$ \\
3 & 128 & 33{,}419 & $2.81 \times 10^{-8}$ & $1.84 \times 10^{-8}$ \\
\midrule
4 & 16  & 875      & $7.46 \times 10^{-7}$ & $4.39 \times 10^{-7}$ \\
4 & 32  & 3{,}275  & $3.99 \times 10^{-7}$ & $2.74 \times 10^{-7}$ \\
4 & 64  & 12{,}683 & $7.12 \times 10^{-8}$ & $5.16 \times 10^{-8}$ \\
4 & 128 & 49{,}931 & $1.90 \times 10^{-8}$ & $2.65 \times 10^{-8}$ \\
\bottomrule
\end{tabular}
\caption{Correction-network capacity ablation on the double-well
$a{=}1.5$ ground state. Mean and standard deviation of the absolute
energy error over 20 independent seeds per configuration. Reference:
$E_0^\mathrm{FD} = 0.80107594$~Ha at $N=1024$.}
\label{tab:ablation_mlp}
\end{table}

The mean error decreases monotonically from $9.0 \times 10^{-7}$~Ha
for the smallest architecture ($2 \times 16$, 331 parameters) to
$1.9 \times 10^{-8}$~Ha for the largest ($4 \times 128$, 49{,}931
parameters), a factor of roughly $50$ over a $150\times$ increase in
model capacity. Even the smallest architecture achieves errors more
than two orders of magnitude below the $10^{-5}$~Ha accuracy claim
of the abstract, so the paper's qualitative conclusions are
architecturally insensitive. The largest architectures saturate near
the $N=1024$ grid-resolution floor identified in
Section~\ref{sec:results:ablation:grid}, as expected. The
$4 \times 128$ configuration is used as the default elsewhere in
the paper.

\section{Perturbation-strength ablation: full results}
\label{app:epsilon_ablation}

Table~\ref{tab:ablation_epsilon} reports the full $\varepsilon$
sweep summarised in
Section~\ref{sec:results:ablation:epsilon}: fifteen log-spaced
values $\{0.005, \allowbreak 0.01, \allowbreak 0.015, \allowbreak
0.02, \allowbreak 0.03, \allowbreak 0.05, \allowbreak 0.075,
\allowbreak 0.1, \allowbreak 0.15, \allowbreak 0.2, \allowbreak
0.25, \allowbreak 0.3, \allowbreak 0.4, \allowbreak 0.5,
\allowbreak 0.7\}$ covering roughly two decades. For each
$\varepsilon$, the $4\times 128$ correction-network ansatz is
trained on the double-well $a=1.5$ ground state with 20 independent
seeds (300 runs total), holding $\varepsilon$ fixed throughout
training. The resulting wavefunctions are projected onto the
Hermite basis at $N_\mathrm{max} = 60$, and the eight Bargmann
zeros nearest the origin are extracted as in
Section~\ref{sec:bargmann}. The table reports two complementary
metrics: the mean absolute energy error
$|E_0^\mathrm{VA} - E_0^\mathrm{FD}|$ (with 20-seed standard
deviation), and the mean absolute position of the leading Bargmann
zero $|z_0|$ together with the standard deviation of the full
zero-set drift vs.\ the $\varepsilon = 0$ symbolic baseline.

\begin{table}[!htbp]
\centering
\begin{tabular}{ccccc}
\toprule
$\varepsilon$ & Mean $|\Delta E|$ (Ha) & Std (Ha) &
$|z_0|_\mathrm{mean}$ & Std($z$-drift) \\
\midrule
0.005 & $2.70 \times 10^{-7}$ & $6.66 \times 10^{-8}$ & 1.6143 & $3.60 \times 10^{-4}$ \\
0.01  & $2.04 \times 10^{-7}$ & $4.93 \times 10^{-8}$ & 1.6143 & $2.77 \times 10^{-4}$ \\
0.015 & $1.65 \times 10^{-7}$ & $3.08 \times 10^{-8}$ & 1.6143 & $2.88 \times 10^{-4}$ \\
0.02  & $1.55 \times 10^{-7}$ & $3.45 \times 10^{-8}$ & 1.6143 & $2.84 \times 10^{-4}$ \\
0.03  & $1.06 \times 10^{-7}$ & $5.26 \times 10^{-8}$ & 1.6143 & $8.58 \times 10^{-4}$ \\
0.05  & $4.36 \times 10^{-8}$ & $4.05 \times 10^{-8}$ & 1.6143 & $9.24 \times 10^{-4}$ \\
0.075 & $1.23 \times 10^{-8}$ & $1.28 \times 10^{-8}$ & 1.6143 & $6.02 \times 10^{-4}$ \\
0.1   & $1.31 \times 10^{-8}$ & $1.31 \times 10^{-8}$ & 1.6143 & $6.34 \times 10^{-4}$ \\
0.15  & $9.35 \times 10^{-9}$ & $6.27 \times 10^{-9}$ & 1.6143 & $4.54 \times 10^{-4}$ \\
0.2   & $1.20 \times 10^{-8}$ & $1.45 \times 10^{-8}$ & 1.6143 & $5.77 \times 10^{-4}$ \\
0.25  & $7.35 \times 10^{-9}$ & $5.68 \times 10^{-9}$ & 1.6143 & $3.62 \times 10^{-4}$ \\
0.3   & $1.53 \times 10^{-8}$ & $1.92 \times 10^{-8}$ & 1.6143 & $6.41 \times 10^{-4}$ \\
0.4   & $1.00 \times 10^{-8}$ & $1.35 \times 10^{-8}$ & 1.6143 & $4.57 \times 10^{-4}$ \\
0.5   & $2.42 \times 10^{-8}$ & $3.72 \times 10^{-8}$ & 1.6143 & $8.45 \times 10^{-4}$ \\
0.7   & $1.19 \times 10^{-8}$ & $1.11 \times 10^{-8}$ & 1.6143 & $5.33 \times 10^{-4}$ \\
\bottomrule
\end{tabular}
\caption{Perturbation-strength ablation on the double-well $a{=}1.5$
ground state ($4\times 128$ correction network, 20 seeds per
$\varepsilon$). Mean absolute energy error and its standard
deviation across seeds (columns 2--3), mean modulus of the leading
Bargmann zero $|z_0|$ (column 4), and standard deviation of the
full zero-set drift vs.\ the $\varepsilon=0$ symbolic baseline
(column 5). Reference: $E_0^\mathrm{FD} = 0.80107594$~Ha.}
\label{tab:ablation_epsilon}
\end{table}

The energy error decreases smoothly from $\sim 3 \times 10^{-7}$~Ha
at $\varepsilon = 0.005$ to below $10^{-8}$~Ha at
$\varepsilon \geq 0.1$, reflecting the increasing expressive
capacity of the ansatz as the correction is allowed greater
amplitude. In contrast, the leading-zero position $|z_0|$ is
pinned at $1.6143$ across all fifteen $\varepsilon$ values to four
decimal places: its seed-to-seed standard deviation is $\sim
10^{-4}$ (reported as zero-drift std in the table), and no
systematic drift with $\varepsilon$ is detectable at this
precision. The energy is refined by $\varepsilon$ by more than an
order of magnitude across the sweep, yet the physical structure
carried by the zero set --- the imaginary-axis location of $z_0$ in
the deep-tunneling ground state --- is invariant.


\bibliographystyle{plainnat}
\bibliography{references}

\begin{thebibliography}{19}
\providecommand{\natexlab}[1]{#1}
\providecommand{\url}[1]{\texttt{#1}}
\expandafter\ifx\csname urlstyle\endcsname\relax
  \providecommand{\doi}[1]{doi: #1}\else
  \providecommand{\doi}{doi: \begingroup \urlstyle{rm}\Url}\fi

\bibitem[Bargmann(1961)]{bargmann1961fock}
V.~Bargmann.
\newblock On a {H}ilbert space of analytic functions and an associated integral
  transform. {P}art {I}.
\newblock \emph{Communications on Pure and Applied Mathematics}, 14\penalty0
  (3):\penalty0 187--214, 1961.
\newblock \doi{10.1002/cpa.3160140303}.

\bibitem[Carleo and Troyer(2017)]{carleo2017science}
Giuseppe Carleo and Matthias Troyer.
\newblock Solving the quantum many-body problem with artificial neural
  networks.
\newblock \emph{Science}, 355\penalty0 (6325):\penalty0 602--606, 2017.
\newblock \doi{10.1126/science.aag2302}.
\newblock arXiv:1606.02318.

\bibitem[Cerf et~al.(2025)Cerf, Wassner, Davis, Arzani, and
  Chabaud]{cerf2025zeros}
Sacha Cerf, Clara Wassner, Jack Davis, Francesco Arzani, and Ulysse Chabaud.
\newblock On the complex zeros of the wavefunction.
\newblock \emph{arXiv preprint arXiv:2507.23468}, 2025.
\newblock \doi{10.48550/arXiv.2507.23468}.

\bibitem[Coleman(1977)]{coleman1977fate}
Sidney Coleman.
\newblock Fate of the false vacuum: Semiclassical theory.
\newblock \emph{Physical Review D}, 15\penalty0 (10):\penalty0 2929--2936,
  1977.
\newblock \doi{10.1103/PhysRevD.15.2929}.

\bibitem[Garg(2000)]{garg2000tunnel}
Anupam Garg.
\newblock Tunnel splittings for one-dimensional potential wells revisited.
\newblock \emph{American Journal of Physics}, 68\penalty0 (5):\penalty0
  430--437, 2000.
\newblock \doi{10.1119/1.19458}.
\newblock arXiv:cond-mat/0003115.

\bibitem[Glauber(1963)]{glauber1963coherent}
Roy~J. Glauber.
\newblock Coherent and incoherent states of the radiation field.
\newblock \emph{Physical Review}, 131\penalty0 (6):\penalty0 2766--2788, 1963.
\newblock \doi{10.1103/PhysRev.131.2766}.

\bibitem[Harris et~al.(2020)Harris, Millman, {van der Walt}, Gommers, Virtanen,
  Cournapeau, Wieser, Taylor, Berg, Smith, Kern, Picus, Hoyer, {van Kerkwijk},
  Brett, Haldane, {del R{\'\i}o}, Wiebe, Peterson, G{\'e}rard-Marchant,
  Sheppard, Reddy, Weckesser, Abbasi, Gohlke, and Oliphant]{harris2020numpy}
Charles~R. Harris, K.~Jarrod Millman, St{\'e}fan~J. {van der Walt}, Ralf
  Gommers, Pauli Virtanen, David Cournapeau, Eric Wieser, Julian Taylor,
  Sebastian Berg, Nathaniel~J. Smith, Robert Kern, Matti Picus, Stephan Hoyer,
  Marten~H. {van Kerkwijk}, Matthew Brett, Allan Haldane, Jaime~Fern{\'a}ndez
  {del R{\'\i}o}, Mark Wiebe, Pearu Peterson, Pierre G{\'e}rard-Marchant, Kevin
  Sheppard, Tyler Reddy, Warren Weckesser, Hameer Abbasi, Christoph Gohlke, and
  Travis~E. Oliphant.
\newblock Array programming with {N}um{P}y.
\newblock \emph{Nature}, 585\penalty0 (7825):\penalty0 357--362, 2020.
\newblock \doi{10.1038/s41586-020-2649-2}.

\bibitem[Hermann et~al.(2020)Hermann, Sch{\"a}tzle, and
  No{\'e}]{hermann2020paulinet}
Jan Hermann, Zeno Sch{\"a}tzle, and Frank No{\'e}.
\newblock Deep-neural-network solution of the electronic {S}chr{\"o}dinger
  equation.
\newblock \emph{Nature Chemistry}, 12\penalty0 (10):\penalty0 891--897, 2020.
\newblock \doi{10.1038/s41557-020-0544-y}.
\newblock arXiv:1909.08423.

\bibitem[Husimi(1940)]{husimi1940formal}
K{\^o}di Husimi.
\newblock Some formal properties of the density matrix.
\newblock \emph{Proceedings of the Physico-Mathematical Society of Japan},
  22\penalty0 (4):\penalty0 264--314, 1940.
\newblock \doi{10.11429/ppmsj1919.22.4_264}.

\bibitem[Janowicz and Zembrzuski(2022)]{janowicz2022bargmann}
Maciej Janowicz and Andrzej Zembrzuski.
\newblock Guessing quantum states from images of their zeros in the complex
  plane.
\newblock \emph{Machine Graphics \& Vision}, 31\penalty0 (3/4):\penalty0
  147--159, 2022.
\newblock \doi{10.22630/MGV.2022.31.3.8}.

\bibitem[Kingma and Ba(2015)]{kingma2015adam}
Diederik~P. Kingma and Jimmy Ba.
\newblock {A}dam: A method for stochastic optimization.
\newblock In \emph{3rd International Conference on Learning Representations
  (ICLR)}, 2015.
\newblock \doi{10.48550/arXiv.1412.6980}.
\newblock arXiv:1412.6980.

\bibitem[Klauder and Skagerstam(1985)]{klauder1985coherent}
John~R. Klauder and Bo-Sture Skagerstam.
\newblock \emph{Coherent States: Applications in Physics and Mathematical
  Physics}.
\newblock World Scientific, Singapore, 1985.
\newblock \doi{10.1142/0096}.

\bibitem[Leboeuf and Voros(1990)]{leboeuf1990zeros}
P.~Leboeuf and A.~Voros.
\newblock Chaos-revealing multiplicative representation of quantum eigenstates.
\newblock \emph{Journal of Physics A: Mathematical and General}, 23\penalty0
  (10):\penalty0 1765--1774, 1990.
\newblock \doi{10.1088/0305-4470/23/10/017}.

\bibitem[Paszke et~al.(2019)Paszke, Gross, Massa, Lerer, Bradbury, Chanan,
  Killeen, Lin, Gimelshein, Antiga, Desmaison, K{\"o}pf, Yang, DeVito, Raison,
  Tejani, Chilamkurthy, Steiner, Fang, Bai, and Chintala]{paszke2019pytorch}
Adam Paszke, Sam Gross, Francisco Massa, Adam Lerer, James Bradbury, Gregory
  Chanan, Trevor Killeen, Zeming Lin, Natalia Gimelshein, Luca Antiga, Alban
  Desmaison, Andreas K{\"o}pf, Edward Yang, Zach DeVito, Martin Raison, Alykhan
  Tejani, Sasank Chilamkurthy, Benoit Steiner, Lu~Fang, Junjie Bai, and Soumith
  Chintala.
\newblock {P}y{T}orch: An imperative style, high-performance deep learning
  library.
\newblock In \emph{Advances in Neural Information Processing Systems},
  volume~32, pages 8024--8035, 2019.
\newblock \doi{10.48550/arXiv.1912.01703}.
\newblock arXiv:1912.01703.

\bibitem[Perelomov(1972)]{perelomov1972coherent}
A.~M. Perelomov.
\newblock Coherent states for arbitrary {L}ie group.
\newblock \emph{Communications in Mathematical Physics}, 26\penalty0
  (3):\penalty0 222--236, 1972.
\newblock \doi{10.1007/BF01645091}.

\bibitem[Pfau et~al.(2020)Pfau, Spencer, Matthews, and
  Foulkes]{pfau2020ferminet}
David Pfau, James~S. Spencer, Alexander G. de~G. Matthews, and W.~M.~C.
  Foulkes.
\newblock Ab initio solution of the many-electron {S}chr{\"o}dinger equation
  with deep neural networks.
\newblock \emph{Physical Review Research}, 2\penalty0 (3):\penalty0 033429,
  2020.
\newblock \doi{10.1103/PhysRevResearch.2.033429}.
\newblock arXiv:1909.02487.

\bibitem[Simon(1970)]{simon1970anharmonic}
Barry Simon.
\newblock Coupling constant analyticity for the anharmonic oscillator.
\newblock \emph{Annals of Physics}, 58\penalty0 (1):\penalty0 76--136, 1970.
\newblock \doi{10.1016/0003-4916(70)90240-X}.

\bibitem[Virtanen et~al.(2020)Virtanen, Gommers, Oliphant, Haberland, Reddy,
  Cournapeau, Burovski, Peterson, Weckesser, Bright, {van der Walt}, Brett,
  Wilson, Millman, Mayorov, Nelson, Jones, Kern, Larson, Carey, Polat, Feng,
  Moore, VanderPlas, Laxalde, Perktold, Cimrman, Henriksen, Quintero, Harris,
  Archibald, Ribeiro, Pedregosa, {van Mulbregt}, and {SciPy 1.0
  Contributors}]{virtanen2020scipy}
Pauli Virtanen, Ralf Gommers, Travis~E. Oliphant, Matt Haberland, Tyler Reddy,
  David Cournapeau, Evgeni Burovski, Pearu Peterson, Warren Weckesser, Jonathan
  Bright, St{\'e}fan~J. {van der Walt}, Matthew Brett, Joshua Wilson, K.~Jarrod
  Millman, Nikolay Mayorov, Andrew R.~J. Nelson, Eric Jones, Robert Kern, Eric
  Larson, C~J Carey, {\.I}lhan Polat, Yu~Feng, Eric~W. Moore, Jake VanderPlas,
  Denis Laxalde, Josef Perktold, Robert Cimrman, Ian Henriksen, E.~A. Quintero,
  Charles~R. Harris, Anne~M. Archibald, Ant{\^o}nio~H. Ribeiro, Fabian
  Pedregosa, Paul {van Mulbregt}, and {SciPy 1.0 Contributors}.
\newblock {S}ci{P}y 1.0: Fundamental algorithms for scientific computing in
  {P}ython.
\newblock \emph{Nature Methods}, 17\penalty0 (3):\penalty0 261--272, 2020.
\newblock \doi{10.1038/s41592-019-0686-2}.

\bibitem[Zinn-Justin and Jentschura(2004)]{zinnjustin2004multiinstanton}
Jean Zinn-Justin and Ulrich~D. Jentschura.
\newblock Multi-instantons and exact results {II}: Specific cases, higher-order
  effects, and numerical calculations.
\newblock \emph{Annals of Physics}, 313\penalty0 (2):\penalty0 269--325, 2004.
\newblock \doi{10.1016/j.aop.2004.04.003}.
\newblock arXiv:quant-ph/0501137.

\end{thebibliography}

\end{document}